\documentclass[acmsmall]{acmart}

\setcopyright{acmlicensed}
\acmJournal{POMACS}
\acmYear{2024} \acmVolume{8} \acmNumber{3}
\acmArticle{33} \acmMonth{12} \acmPrice{15.00}
\acmDOI{10.1145/3700432}

\received{August 2024}
\received[revised]{September 2024}
\received[accepted]{October 2024}

\usepackage{tikz}
\usepackage{amsmath}

\usepackage{amssymb}
\usepackage{amsfonts}
\usepackage{textcomp}
\usepackage{xcolor}
\usepackage{url}
\usepackage{afterpage}
\usepackage{multirow}

\usepackage{algorithm}
\usepackage{algorithmic}
\usepackage{graphicx}
\usepackage{subfigure} 
\usepackage{fancybox}
\usepackage{xspace}
\usepackage{threeparttable}
\usepackage{multirow}
\usepackage{enumitem}
\usepackage{seqsplit}
\usepackage{tcolorbox}
\tcbuselibrary{breakable}

\author{Yanhui Guo}
\affiliation{%
\institution{Beijing University of Posts and Telecommunications}
\city{Beijing}
\country{China}
}

\author{Dong Wang}
\affiliation{%
\institution{Beijing University of Posts and Telecommunications}
\city{Beijing}
\country{China}
}

\author{Liu Wang}
\affiliation{%
\institution{Beijing University of Posts and Telecommunications}
\city{Beijing}
\country{China}
}

\author{Yongsheng Fang}
\affiliation{%
\institution{Beijing University of Posts and Telecommunications}
\city{Beijing}
\country{China}
}

\author{Chao Wang}
\authornotemark[2]
\affiliation{%
  \institution{Huazhong University of Science and Technology}
  \city{Wuhan}
  \state{Hubei}
  \country{China}}

\author{Minghui Yang}
\affiliation{%
  \institution{OPPO}
  \city{Shenzhen}
  \state{Guangdong}
  \country{China}}

\author{Tianming Liu}
\authornote{Corresponding authors: Haoyu Wang (haoyuwang@hust.edu.cn) and Tianming Liu (Tianming.Liu@monash.edu).}
\authornotemark[2]
\affiliation{%
  \institution{Monash University}
  \city{Melbourne}
  \state{Victoria}
  \country{Australia}
  }
\affiliation{%
  \institution{Huazhong University of Science and Technology}
  \city{Wuhan}
  \state{Hubei}
  \country{China}
  }

\author{Haoyu Wang}
\authornotemark[1]
\authornote{The full name of the authors' affiliation with Huazhong University of Science and Technology is: Hubei Key Laboratory of Distributed System Security, Hubei Engineering Research Center on Big Data Security, School of Cyber Science and Engineering, Huazhong University of Science and Technology.}
\affiliation{%
  \institution{Huazhong University of Science and Technology}
  \city{Wuhan}
  \state{Hubei}
  \country{China}}

\renewcommand{\shortauthors}{Yanhui Guo et al.}



\usepackage{hyperref}
\AtEndPreamble{
	\usepackage{hyperref}
	\hypersetup{
		colorlinks = true,
		linkcolor = brown,
		anchorcolor = brown,
		citecolor = brown,
		filecolor = brown,
		urlcolor = brown
	}
}

\newcommand{\answer}[2] {
	\begin{tcolorbox}[breakable, boxrule=1pt,left=1pt,right=1pt,top=2pt,bottom=2pt]
		\textbf{Answer to RQ#1:} #2
	\end{tcolorbox}
}

\definecolor{customblue}{HTML}{006ca6}
\definecolor{customgreen}{HTML}{009264}
\definecolor{custombrown}{HTML}{ff3d00}

\begin{document}
\title{Beyond App Markets: Demystifying Underground Mobile App Distribution Via Telegram}

\begin{abstract}
The thriving mobile app ecosystem encompasses a wide range of functionalities. However, within this ecosystem, a subset of apps provides illicit services such as gambling and pornography to pursue economic gains, collectively referred to as "underground economy apps". While previous studies have examined these apps' characteristics and identification methods, investigations into their distribution via platforms beyond app markets (like Telegram) remain scarce, which has emerged as a crucial channel for underground activities and cybercrime due to the robust encryption and user anonymity.

This study provides the first comprehensive exploration of the underground mobile app ecosystem on Telegram. Overcoming the complexities of the Telegram environment, we build a novel dataset and analyze the prevalence, promotional strategies, and characteristics of these apps.
Our findings reveal the significant prevalence of these apps on Telegram, with the total sum of subscription user numbers across channels promoting these apps equivalent to 1\% of Telegram's user base\footnote{This percentage is calculated using total subscription user numbers without accounting for potential user overlaps, please refer to \S\ref{sec:scopenum} for a detailed explanation and discussion of implications.}.
We find these apps primarily cater to gambling and pornography services.
We uncover sophisticated promotional strategies involving complex networks of apps, websites, users, and channels, and identify significant gaps in Telegram's content moderation capabilities. Our analysis also exposes the misuse of iOS features for app distribution and the prevalence of malicious behaviors in these apps.
This research not only enhances our understanding of the underground app ecosystem but also provides valuable insights for developing effective regulatory measures and protecting users from potential risks associated with these covert operations. Our findings provide implications for platform regulators, app market operators, law enforcement agencies, and cybersecurity professionals in combating the proliferation of underground apps on encrypted messaging platforms.
\end{abstract}

\maketitle

\section{Introduction}
\label{sec:intro}
Mobile apps have developed into an extensive and complex ecosystem, covering various fields ranging from entertainment to education. 
However, within this process, a portion of apps serve the underground economy, driven by the pursuit of significant economic gains, giving rise to so-called "underground economy apps"~\cite{chen2023deuedroid}.
These apps provide non-compliant services in sensitive areas such as gambling and pornography, violating or skirting legal regulations, posing significant challenges to digital security and law enforcement, thereby attracting wide attention from academia and industry. 
Current research on underground apps has covered the promotional websites of these apps~\cite{han2023measurement,gao2021demystifying,chen2024underground}, the characteristics of underground apps~\cite{gao2021demystifying, hu2022measurement, hong2022analyzing, chen2021lifting, hu2019dating}, and methods for identifying these apps~\cite{chen2023deuedroid,zhao2024no}.

Telegram is a widely popular instant messaging platform, whose encryption features and commitment to user anonymity have made it a preferred channel for the dissemination of underground economy and cybercrime content~\cite{tele10Guards, telecloudsek}. While investigations into Telegram's underground activities, including conspiracy channels~\cite{imperati2023conspiracy}, identification of clone and fake channels~\cite{la2021uncovering}, and analysis of misinformation distribution~\cite{la2023sa}, have emerged, studies examining the underground mobile app ecosystem within Telegram are still missing, which is crucial for developing targeted regulatory strategies and protecting users from potential risks.

This research is dedicated to exploring the ecosystem of underground mobile apps within the Telegram environment (\textbf{TUApps} for short). 
We aim to provide comprehensive insights into the TUApps ecosystem, including the entities involved, supporting networks, operational methods, and their impact on users and the digital marketplace.
To achieve these objectives, our research is structured around three key research questions (\S\ref{sec:rq}):
What is the prevalence and accessibility of underground mobile apps on
Telegram?
What are the strategies for promoting these apps on Telegram? And, what are the development and behavioral characteristics of these apps?

The first challenge in this research was to obtain a set of underground apps from Telegram. 
Telegram has a large and widely distributed user base, making the filtering and identification of underground activities inefficient and resource-intensive. 
Our breakthrough came upon finding that Telegram hosts bots equipped with search capabilities, allowing us to efficiently trawl through channels using specified keywords.
To enhance our dataset's breadth and depth, we expanded our channel list by incorporating popular channels and employing a snowballing technique (\S\ref{keywords}).
We then extracted website URLs promoting underground apps from the message histories of these channels (\S\ref{sec:crawl-message}), and retrieved the apps from these promotional websites with a semi-automated crawling method (\S\ref{sec:semicrawl}). 
This process established a robust foundation for our analysis of the underground app ecosystem on Telegram.
Subsequently, we conducted an in-depth exploration of the TUApps ecosystem, guided by our three research questions (\S\ref{sec:rq1sec},\ref{appPromtionSection},\ref{sec:rq3sec}). Our findings underscore the urgent need for enhanced regulation of underground apps on Telegram and reveal several critical implications. Our research also illuminates potential avenues for mitigation, offering valuable insights not only to platform regulators but also to a broader range of stakeholders including app market operators and law enforcement agencies (\S\ref{sec:miti}).

In summary, the contributions of this paper are as follows:
\begin{itemize}
    \item We conducted a comprehensive study on the ecosystem of underground apps on Telegram, which is the first of its kind. We contributed a novel dataset\footnote{Our code and dataset are available at \url{https://github.com/security-pride/TUApps}. Kindly refer to \S\ref{ava} for more details on availability.} comprising over 200 million messages, 10,000 apps downloaded, and 1,000 unique apps. Our research investigated multiple aspects of these apps, including their prevalence, promotional strategies, and characteristics.\looseness=-1
    \item We uncovered the significant presence of TUApps on Telegram, nominally reaching 1\% of the platform's user base. These apps primarily cater to gambling and pornography services. In terms of availability, they are distributed mainly through APK, IPA, and Web Clip files rather than being available on official app stores. Notably, we identified the misuse of iOS features like TestFlight, Enterprise Signing, and Web Clips for TUApp distribution, revealing vulnerabilities in the iOS ecosystem.
    \item We unveiled sophisticated promotional strategies of TUApps, detailing a complex promotional framework comprising interlinked relationships among apps, promotional websites, users, and channels. Our analysis revealed that promotional websites use diverse hosting strategies while maintaining structural URL similarities, offering potential detection avenues. We identified significant gaps in Telegram's ability to detect potentially harmful content, with only 0.28\% of promotional channels flagged as scams.
    We also revealed the presence of dedicated app promotion teams active on Telegram.
    \item We analyzed the development and behavioral characteristics of TUApps, revealing that a quarter of them are malicious, posing substantial security risks. We found their preference for cross-platform development frameworks such as Flutter for ease of development. In terms of payment methods, TUApps frequently utilize fourth-party payment platforms to evade regulatory scrutiny.
\end{itemize}
\section{Background}
\subsection{Telegram}

Telegram is a cross-platform encrypted instant messaging service with over 800 million users~\cite{teleUserNum} that originated in 2013, and is known for its security, speed, and rich features. It allows users to communicate one-on-one through text, images, videos, and files. Additionally, Telegram provides channel and group functionalities, enabling users to interact within broader social circles. 

With a maximum capacity of 200,000 members, groups support interactive communication among multiple users, whereas channels are designed for broadcast communication from administrators to an unlimited audience. Telegram allows these channels and groups to be set as public (i.e., can be found with Telegram search, and every user can access historical messages and join the group) or private (i.e., only members can see the posts, and users need an invite link to join it). Despite the differentiation between channels and groups, this distinction is inconsequential for our research. Throughout this research, we employ the term "channel" to encompass both channels and groups.

Due to the anonymity and the convenience of Telegram communication, it has become a natural platform for the dissemination of underground economy and cybercrime, such as money laundering~\cite{mule}, financial fraud~\cite{telecrime2}, revenge porn~\cite{telePron}, and the trade of personal information~\cite{telehacknews}. It has been pointed out by many news media and reports~\cite{tele10Guards, telecloudsek, telecrime, telecrime2} that Telegram has evolved into a "cybercrime ecosystem" resembling dark web forums, attracting criminals such as cyber thieves and hackers.
To boost platform security and user trust, Telegram has introduced the "verified" and the "scam" marks to channels and accounts to alert users about potential fraudulent activities.

\subsection{Bots in Telegram}
\label{bot}
Telegram offers the Bot API~\cite{TelegramAPI} to developers, enabling them to create bots with diverse functionalities tailored to specific requirements. These functionalities encompass information retrieval, automation, and productivity tools, among others.
One notable type of bot is the "search bot", which utilizes user-provided keywords to conduct channel searches and curate a selection of channels matching those keywords. This empowers users to discover and explore channels that closely align with their interests and preferences.

Based on our preliminary investigation, we discovered that search bots are frequently utilized by underground actors for advertising and promotion purposes. The working mechanism of these search bots revolves around keywords, which are sold by the bot owners for profit. When users search for a keyword that has been purchased, the buyer's content is strategically displayed at a higher rank or in a pinned position, ensuring maximum visibility.

\begin{figure}[htb]
	
	\begin{minipage}{0.4\linewidth}
		\vspace{3pt}
		\centerline{\includegraphics[width=\textwidth]{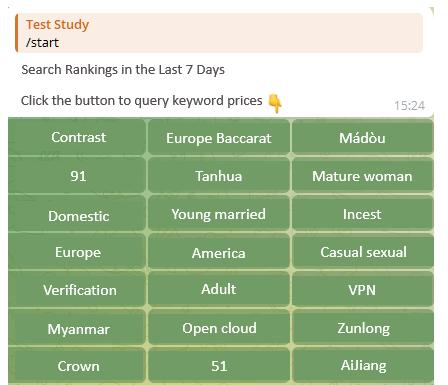}}
		\centerline{(a)}
	\end{minipage}
	\begin{minipage}{0.4\linewidth}
		\vspace{3pt}
		\centerline{\includegraphics[width=\textwidth]{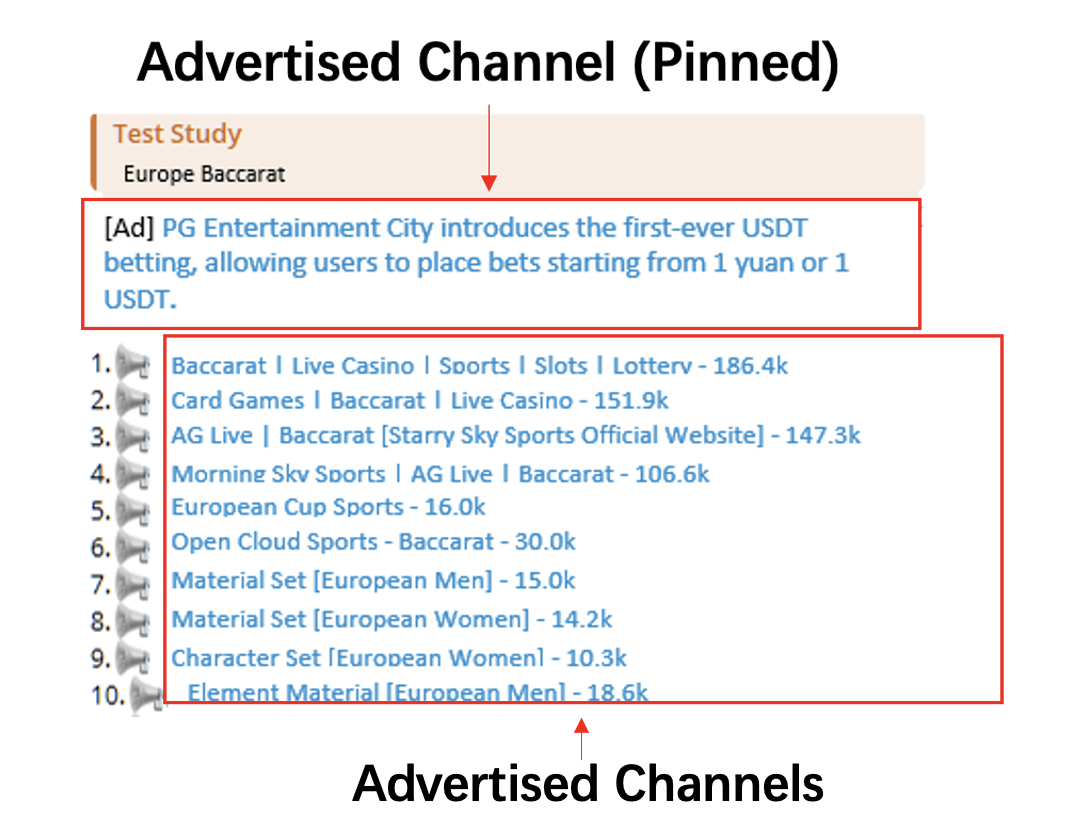}}
		\centerline{(b)}
	\end{minipage}
 
	\caption{Keywords List in Search Bot and Search Results for "Europe Baccarat"}
    \label{KeywordsList}
\end{figure}

A prime example of such a search bot is "Search Groups Chinese Search"~\cite{TGbaiduCN}, which boasts a user base of over 64,000 Telegram users. As illustrated in  \autoref{KeywordsList}(a), the bot offers a wide array of keywords for sale, with the majority pertaining to the underground economy.
When a user searches for one of these keywords to find related channels, the search bot returns results that prioritize the buyer's promotional content. For instance, as depicted in \autoref{KeywordsList}(b), a search for gambling content related to "Europe Baccarat\footnote{A popular card game in casinos.}" yields results where the pinned item and the top 10 search results all feature advertised channels associated with "Europe Baccarat" gambling activities.
The prices of these ad slots ranged from \$660 to \$750.

\subsection{App Installation Package}
Android and iOS, the two dominant mobile operating systems, employ distinct app installation package formats and distribution mechanisms.

On the Android platform, apps are packaged in the APK (Android Package Kit) format. The open nature of the Android ecosystem allows for app distribution through multiple channels, including the official Google Play Store, third-party marketplaces, and direct APK file downloads. While Android systems by default restrict the installation of APKs from unknown sources, users can easily modify this setting. This flexibility inadvertently facilitates underground app developers in guiding users to bypass security restrictions.

Conversely, the iOS platform uses the IPA (iOS App Store Package) format for app installation. iOS enforces stricter controls, generally prohibiting the installation of IPA files from unauthorized sources. Ordinary users are typically limited to downloading and installing apps exclusively through the official Apple App Store. However, iOS does provide alternative distribution methods, ostensibly for specific purposes such as app testing. These include TestFlight, Enterprise Signing, and Web Clip. Our research reveals that these mechanisms are being exploited to distribute underground apps, a topic we will explore in greater detail in \S\ref{distributionform}.

\section{Study Design}

\subsection{Research Questions}
\label{sec:rq}
Our study is driven by the following research questions (RQs):

\begin{itemize}
    \item \textbf{RQ1: What is the prevalence and accessibility of underground mobile apps on Telegram?} This question aims to explore the types of underground economy these apps serve, their distribution scope on Telegram channels, their availability (or lack thereof) on official app stores, and the various distribution methods employed to make these apps accessible to users, with particular attention to the unique restrictions on the iOS platform.
    \item \textbf{RQ2: What are the strategies for promoting underground mobile apps on Telegram?} This question examines the intricate promotional networks that facilitate the spread of underground mobile apps within Telegram. Focusing on the roles of users, channels, and websites, we aim to unravel how these entities collaborate to amplify the visibility and accessibility of these apps. 
    \item \textbf{RQ3: What are the characteristics of the underground mobile apps promoted on Telegram?} With these apps typically engaging in illicit activities or distributing prohibited content, and opting for Telegram to sidestep detection and scrutiny, it becomes imperative to investigate their attributes. This includes an examination of their development and behavior characteristics, to acquire a thorough understanding of how these underground apps function.
\end{itemize}

\subsection{Dataset Collection}
\label{sec:datacollect}
\begin{figure*}[!t]
  \centering
  \includegraphics[width=\textwidth]{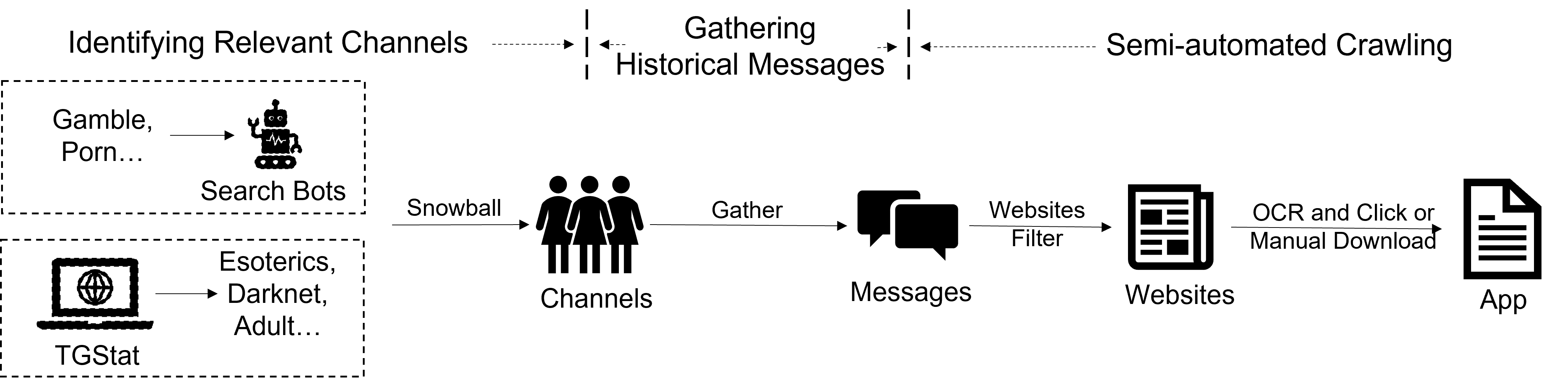}
  \caption{Dataset Collection}
  \label{datasetCollection}
\end{figure*}

To answer our research questions, we collect underground apps promoted within public Telegram channels that could be associated with underground activities.
\autoref{datasetCollection} shows our data collection process, which consists of three steps: (1) identifying relevant channels in Telegram, (2) obtaining historical messages from channels, and (3) semi-automated crawling of apps.

\subsubsection{Identifying Relevant Channels}
\label{keywords}
To sample relevant channels, we leverage Telegram's ``search'' function and the snowballing-based technique to retrieve a set of channels related to underground activities. Our approach was threefold: 

(1) 
As previously mentioned, search bots are often employed to promote underground content. We hence leverage search bots to kick-start our channel acquisition process.
To identify effective search bots, we refer to the chart of top-100 bots on TelegramChannels~\cite{TelegramChannels}, a popular third-party channel and bot ranking platform. 
We supplemented this selection with 8 additional bots discovered through our daily interactions on the platform.
As introduced in \S\ref{bot}, these bots offer keywords related to underground activities for sale.
We thus curated a collection of nearly 300 keywords associated with underground activities in both English and Chinese from the purchasable keywords listed in these search bots\footnote{The bot and keyword lists are available on our website.}.
These keywords were employed in our interactions with these bots to curate a contextually relevant collection of channels. This yields an initial list of 15,796 channels.

(2) Considering that underground apps may also be promoted through regular channels, we enhance our channel collecting strategy. Specifically, we rely on the rankings from a third-party website, TGStat~\cite{tgstat}, that provides extensive statistics and analytics for Telegram channels and groups\footnote{TGStat is also frequently utilized by previous research on Telegram, such as ~\cite{la2021uncovering,la2023tgdataset,tikhomirova2021community}. We use a different website for bot ranking because TGStat provides no such service.}. TGStat categorizes Telegram channels into over 40 categories, including regular categories such as News and Tech, as well as categories related to underground economy such as Darknet and Adult. We extract the top 100 channels with the most subscribers from each category listed on TGStat. This yields another list of 4,700 popular channels covering various categories.

(3) With the above two lists of channels as `seed channels', we next incrementally expand our channel list with potentially relevant public channels whose messages have been forwarded to existing seed channels. Such a snowballing approach involves crawling the historical messages of these seed channels (as detailed in \S\ref{sec:crawl-message}), parsing the message content, and extracting channel URLs from which messages had been forwarded. This effort yielded an extended list of 71,332 public Telegram channels for our research.

\subsubsection{Gathering Historical Messages from Channels}
\label{sec:crawl-message}
Following the compilation of a list of potentially relevant channels, we next crawled historical messages from these channels.
The historical messages were collected using Telegram's open API and Telethon Python library~~\cite{telethon}.
In principle, all messages ever posted in a public channel are available. However, considering the time overhead of the crawling effort and the timeliness of the messages (web links contained in earlier messages are likely to be dead), we limited our data scraping to a maximum of 100,000 messages from each channel.
In total, we collected over 200 million messages.
This extensive message dataset allows us to explore the dissemination of underground apps within the Telegram ecosystem.

\subsubsection{Semi-automated Crawling of Apps}
\label{sec:semicrawl}
Our goal is to find underground mobile apps circulating in messages on the Telegram platform. Typically, these apps are promoted through web pages or download links. We therefore extract website URLs from the messages using regular expression matching. Due to the substantial volume of URLs (over 8 million) and the presence of irrelevant websites, we implement an additional filter to identify key URLs. Specifically, we filter by the prevalence of the promoted URLs, i.e., the number of channels promoting them, focusing on URLs that are mentioned in more than 20 different channels. 
As a result, we obtain a total of 21,147 website URLs being promoted in over 20 channels.
These selected URLs account for 65.4\% of the entire promotion frequency of the 8 million URLs, offering a concentrated view of the TUApps landscape while ensuring the process remains streamlined and effective.

We then develop a crawler to retrieve the mobile apps linked to each of the selected website URLs (if available). The scraper utilized an open-source OCR toolkit PaddleOCR~\cite{PaddleOCR}, and the open-source webdriver Selenium~\cite{Selenium}, which allows one to control a browser programmatically.
For each website, if accessible, we initiated the process by capturing a screenshot of the webpage. Subsequently, we employ PaddleOCR to recognize the text within the screenshot. If any of the identified text matches keywords associated with app downloads such as `Download', `Android', or `iOS', we determine the text's location and use Selenium to automatically trigger clicks on these locations to initiate app downloads.
To ensure the accuracy of the collection, we further performed a manual examination of the screenshots from the accessed websites. For downloaded apps, we check the screenshots to verify their relevance to underground economy. For cases where websites failed to download automatically, we examined their screenshots to identify any contained apps, and if present, we manually downloaded the respective packages to ensure the inclusion of relevant apps that might have evaded the automated process.

\subsection{Dataset Overview}
\label{datasetOverview}
Our dataset collection process lasted for 6 months from September 2023 to February 2024. The dataset overview is presented in \autoref{staticApp}. In total, we collected 71,332 potentially relevant public channels, from which we obtained over 200 million historical messages and extracted a total of 8,283,875 website URLs from the messages.
Of these, 21,147 URLs were promoted by more than 20 channels, which we visited in turn to retrieve their linked mobile apps. 
As a result, 5,692 websites successfully executed downloads, from which we collected 5,692 Android installation files and 4,968 iOS installation files\footnote{Typically, Android and iOS apps are released simultaneously, meaning that on a website, users can download both the Android and iOS versions. However, some websites only provide the Android version.}, corresponding to 557 unique Android apps and 517 unique iOS apps after de-duplication.
These apps were derived from 2,333,830 messages promoted in 6,137 channels, posted by 7,957 Telegram users.

\begin{table}[htb]
  \centering
  \caption{Overview of the Dataset}
  \small
  \begin{tabular}{cc}
    \hline
    Item & Count \\
    \hline
    Public Channels & 71,332 \\
    Messages & Over 200 million \\
    Total Websites & 8,283,875 \\
    Android Apps Downloaded & 5,692 \\
    iOS Apps  Downloaded & 4,968 \\
    Unique Android Apps & 557 \\
    Unique iOS Apps & 517 \\
    \hline
  \end{tabular}
  \label{staticApp}
\end{table}

\section{Prevalence and Accessibility of TUApps}
\label{sec:rq1sec}
This section examines the prevalence and accessibility of TUApps. We explore the types of underground economy these apps serve, analyze their distribution scope across Telegram channels, investigate their presence (or absence) on public app stores, and detail the various forms employed for their distribution.

\subsection{Classification of TUApps}

To understand the types of underground economy associated with the apps in our dataset, we conducted a manual classification\footnote{Two authors independently performed the classification and then convened to compare and discuss the results.} of the 557 apps.

As shown in \autoref{appType}, the apps were categorized into five distinct groups: pornography, gambling, pirated software, virtual private network (VPN), and cryptocurrency trading platforms.
Notably, pornography and gambling apps constituted the overwhelming majority of the dataset, accounting for over 95\% of the total apps.

\begin{table}[htb]
  \centering
  \small
  \caption{Classification of TUApps}
  \begin{tabular}{ccc}
    \hline
    Types of Underground Economy & \# Count & \% Percent\\
    \hline
    Pornography & 314 & 56.4\\
    Gambling & 228 & 40.9\\
    Pirated Software & 10 & 1.8\\
    Virtual Private Network (VPN) & 3 & 0.5\\
    Cryptocurrency Trading Platform & 2 & 0.3\\
    \hline
  \end{tabular}
  \label{appType}
\end{table}

\subsection{Distribution Scope of TUApps}
\label{sec:scopenum}

To assess the prevalence of TUApps on Telegram, we analyzed their distribution scope, focusing on the number of channels promoting each app and the total promotional messages across these channels. Our analysis revealed a significant presence, with each app, on average, being promoted 4,190 times across 37.17 channels.
Examining the 6,137 channels involved in TUApp promotion, we found an average of 1,409 members per channel, with a total subscription base of 8,647,033 members across all channels. \textbf{This substantial reach, before accounting for potential overlaps, constitutes 1.08\% of Telegram's reported 800 million users} as of 2023~\cite{teleUserNum}.
The most widely distributed app in our dataset, a gambling app called "Zunlong Kai Shi",  was advertised across 216 distinct channels to 217,512 users, with an astonishing 357,951 promotional messages in total.
It is important to note that, due to ethical concerns, our data collection methodology recorded only channel membership numbers without identifying individual users. This approach likely leads to an overestimation of unique individuals exposed to TUApps, as users may be members of multiple channels, suggesting that the actual number of unique individuals is lower than the total sum of 8,647,033 members. However, it is crucial to recognize that our data collection process, as detailed in \S\ref{sec:datacollect}, captures only a partial snapshot of the underground app's presence in the Telegram ecosystem. Our method, which included crawling relevant channels and messages, and sampling URLs occurring in more than 20 different channels, while effective in offering a concentrated view of the TUApps landscape, cannot encompass the entire scope of underground app activity on Telegram. Consequently, while user overlap may inflate the total subscription numbers, the true reach of underground apps on Telegram may still be underestimated.

These findings underscore the intensive and wide-ranging promotional efforts undertaken for TUApps on Telegram, while also highlighting their prevalence on the platform.
The broad distribution of promotional activities across multiple channels suggests that app developers and promoters employ diverse strategies to maximize their reach to target audiences. This approach is likely driven by the intense competition within the TUApps market on Telegram. We further investigate their promotional strategies in \S\ref{appPromtionSection}.

\subsection{Public Availability of TUApps}
We investigated the public availability of the collected apps in our study by searching for their presence in mainstream app markets.

For Android apps, we first used Androguard~\cite{androguard} to extract the package name of each app in our dataset. We then searched for these package names in the official Android marketplace, Google Play~\cite{goole}, and popular third-party app markets such as Xiaomi~\cite{xiaomi} and Huawei~\cite{huawei} App Store. None of the package names from our dataset were found to be indexed in these app markets.

For iOS apps, we extracted each app's name and unique bundle identifier (bundleId) from the \textit{Info.plist} file's \texttt{\seqsplit{CFBundleDisplayName}} and \texttt{\seqsplit{CFBundleIdentifier}} fields in IPA files, and the \texttt{\seqsplit{PayloadDisplayName}} field in iOS Web Clips (elaborated in \S\ref{distributionform}). We then systematically searched for the extracted app names using the iTunes Search API~\cite{iTunesAPI}. For each app, we retrieved the bundleId from the meta-information in its search results for accurate app matching.

\begin{figure}[htb]
	
	\begin{minipage}{0.2\linewidth}
		\vspace{3pt}
		\centerline{\includegraphics[width=\textwidth]{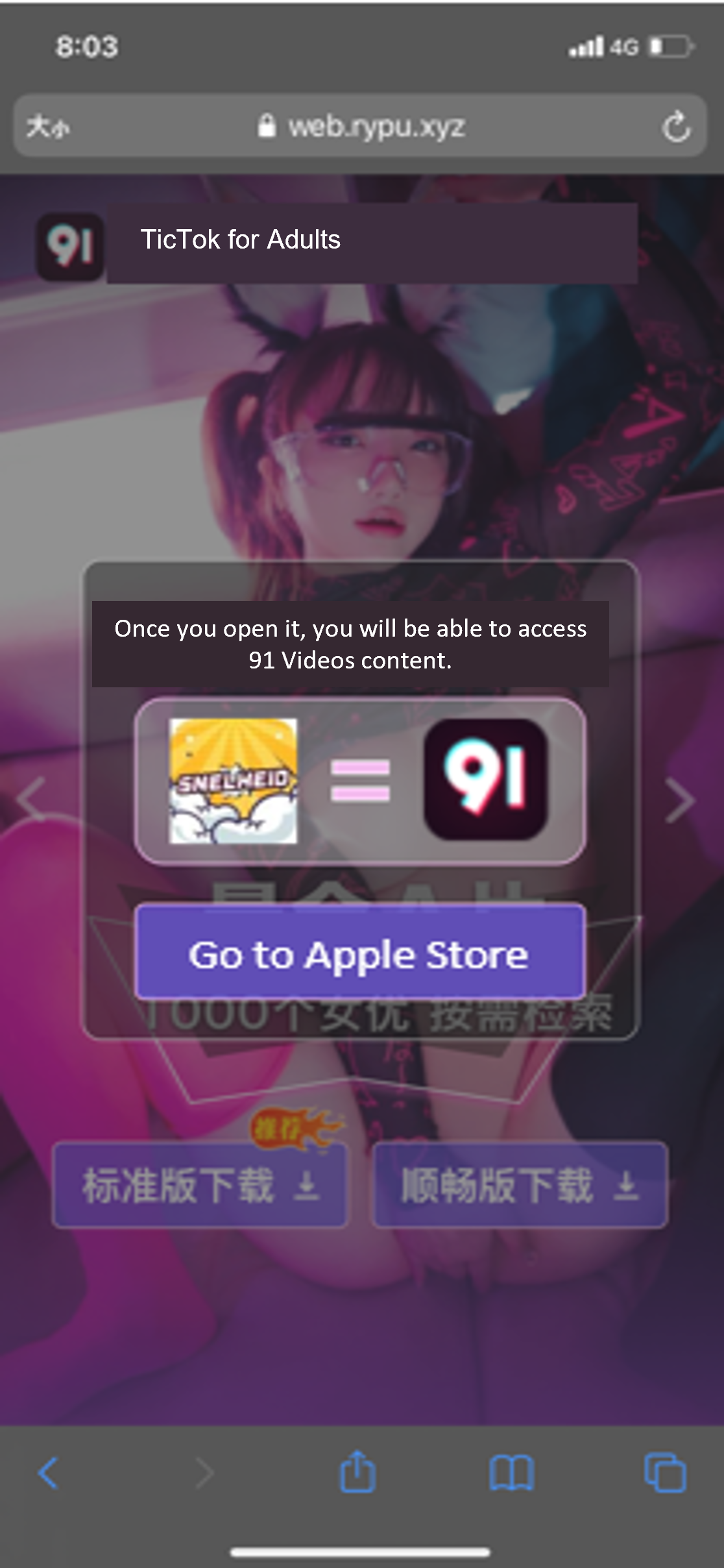}}
        \centerline{(a)}
	\end{minipage}
	\begin{minipage}{0.2\linewidth}
		\vspace{3pt}
		\centerline{\includegraphics[width=\textwidth]{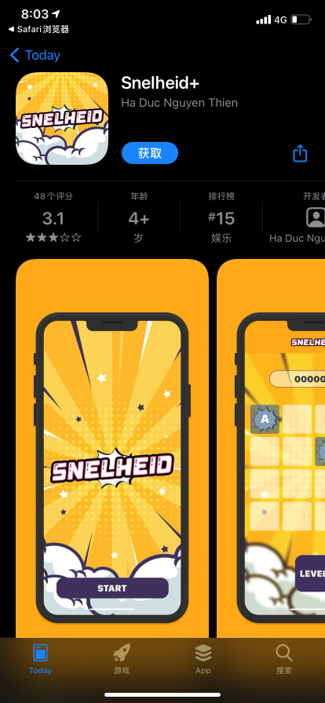}}
        \centerline{(b)}
	\end{minipage}
	\begin{minipage}{0.2\linewidth}
		\vspace{3pt}
		\centerline{\includegraphics[width=\textwidth]{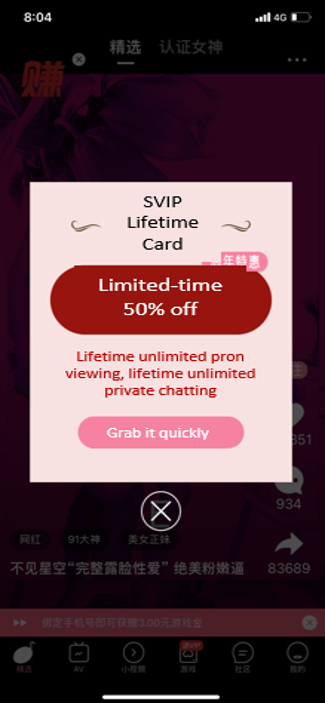}}
        \centerline{(c)}
	\end{minipage}
 
	\caption{Underground App in AppStore}
	\label{UndergroundAppinAppStore}
\end{figure}

Interestingly, we found 12 matches among the iOS apps. Upon closer examination, we discovered that these 12 apps were promoted on Telegram with official Apple App Store links, and they were underground apps disguised as legitimate apps on the App Store.
A most recent research~\cite{zhao2024no} also referred to such apps as "mask apps".
\autoref{UndergroundAppinAppStore}(a) shows an example of one such app being promoted on a website within Telegram. The app, which leads users to the App Store for downloads, is actually a pornographic app masquerading as a regular app. 
Instead of downloading the app under its original name "TicTok for Adults", users are directed to download a game app named "Snelheid+", as shown in \autoref{UndergroundAppinAppStore}(b). The app's icon and description on the App Store appear innocuous. However, upon downloading and opening the app, it is revealed to be filled with pornographic content, as depicted in \autoref{UndergroundAppinAppStore}(c). It is worth noting that by the time of writing, all 12 apps were no longer available on the Apple App Store, likely due to Apple's app vetting process.

Our findings reveal that despite Apple's stringent app vetting procedures, underground actors still find ways to promote their apps using the official App Store. However, such cases are rare on iOS (12 out of 517), and we found no similar instances on Android. This disparity can be attributed to the ease with which Android APKs can be promoted and installed outside of official app markets.

\subsection{Distribution Forms of TUApps}
\label{distributionform}

In terms of distribution forms, Android TUApps are packaged as APK files, while iOS TUApps are distributed as IPA files and Web Clips~\cite{applewebclip}.
Due to the iOS restriction on direct IPA file installation, the IPA files are distributed through alternative methods such as Enterprise Signing~\cite{AppleDeveloperEnterprise} and TestFlight invitations~\cite{testflight}. As shown in \autoref{iOSType}, of the 517 unique iOS apps in our collection, 497 were available as Web Clips, while the remaining 20 IPA files were distributed using Enterprise Signing (16) and TestFlight invitations (4). We elaborate on these three distribution forms below:

\begin{table}[htb]
  \centering
  \small
  \caption{Distribution Forms of iOS TUApps}
  \begin{tabular}{cccc}
    \hline
    Forms & \# Unique App Count & \# Samples Count & \% Samples Percent\\
    \hline
    TestFlight Invitation & 4 & 316 & 6.4 \\
    Enterprise Signing & 16 & 805 & 16.2 \\
    Web Clip Installation & 497 & 3,847 & 77.4 \\
    \hline
  \end{tabular}
  \label{iOSType}
\end{table}

\noindent\textbf{TestFlight Invitation.} Our dataset included 316 samples (6.4\% of 4,968 downloaded iOS samples) distributed via TestFlight. This Apple-provided platform for app testing allows developers to invite up to 10,000 external testers through public links, with apps remaining valid for 90 days. TestFlight's relatively lenient review standards, compared to the App Store, facilitate the distribution of certain underground apps. Users only need to install TestFlight and follow the invitation link to access these apps. Some distributors also share App Store accounts to circumvent tester limitations.

\noindent\textbf{Enterprise Signing.} We identified 805 samples (16.2\%) distributed through Enterprise Signing. This method allows companies to deploy proprietary apps directly to iOS devices in a controlled and customized manner, bypassing the App Store and Apple's standard review process. A key feature of Enterprise Signing is the absence of restrictions on expiration dates or the number of users, making it an attractive option for disseminating underground apps.
However, it is important to note that while both TestFlight and Enterprise Signing undergo less stringent vetting compared to App Store submissions, they remain within Apple's regulatory purview. Apple has been known to ban accounts misusing these methods for illegitimate purposes~\cite{applesideloading}. This likely contributes to the lower prevalence of these distribution methods in our dataset compared to Web Clips.

\noindent\textbf{Web Clip Installation.} The majority of iOS TUApp samples in our dataset (3,847, 77.4\%) were distributed as Web Clips. This iOS feature enables users to add customized shortcut icons to their device's home screen, providing easy access to specific web pages or settings~\cite{applewebclip}.
The process of setting up Web Clips is significantly less complex than guiding users through TestFlight invitations or Enterprise Signing procedures. Despite Web Clips offering only a shortcut to a specific web page rather than a full app experience, most underground actors prefer this method over distributing IPA files on iOS platforms. This preference stems from Web Clips' lack of regulation, as they fall outside Apple's app review process, their affordability due to not requiring a costly personal or enterprise Apple developer account, and their user-friendly nature, demanding minimal effort from users during installation. These advantages have made Web Clips the favored route for disseminating a significant volume of iOS TUApps. This finding aligns with previous research on the prevalence of Web Clips in the underground economy~\cite{hu2022measurement}, underscoring the method's effectiveness in circumventing traditional app distribution channels.

The diversification of these distribution strategies demonstrates that despite iOS's stringent security measures, underground actors continue to discover and exploit loopholes for content distribution. This ongoing cat-and-mouse game underscores the pressing need for more robust regulatory measures and enhanced vetting processes to mitigate abuse of these distribution channels.

\answer{1}{Our analysis reveals that TUApps predominantly cater to pornography and gambling services. These apps exhibit a significant presence on Telegram, with each app being distributed across an average of 37.17 channels in 4,190 messages, and collectively reaching 1\% of Telegram's user base at face value. While official app stores generally do not list these apps, we observed rare instances of their temporary presence on the Apple App Store, which were swiftly removed upon detection. On iOS, the distribution landscape is diverse: Web Clips emerge as the primary distribution method, supplemented by instances of TestFlight and Enterprise Signing abuse for IPA distribution.}
\section{TUApps Promotion in Telegram}
\label{appPromtionSection}
This section examines the multifaceted promotional ecosystem of underground apps on Telegram through three key aspects:
First, we analyze the promotional strategies employed, revealing the techniques used to maximize app visibility and user engagement. Next, we investigate the existence and structure of organized promotion teams, shedding light on the collaborative efforts behind app distribution. Finally, we examine the characteristics of promotional websites, including their longevity and infrastructure, to understand their role in supporting TUApp distribution.

\subsection{Promotion Strategy}

\begin{figure*}[htb]
  \centering
  \includegraphics[width=\textwidth]{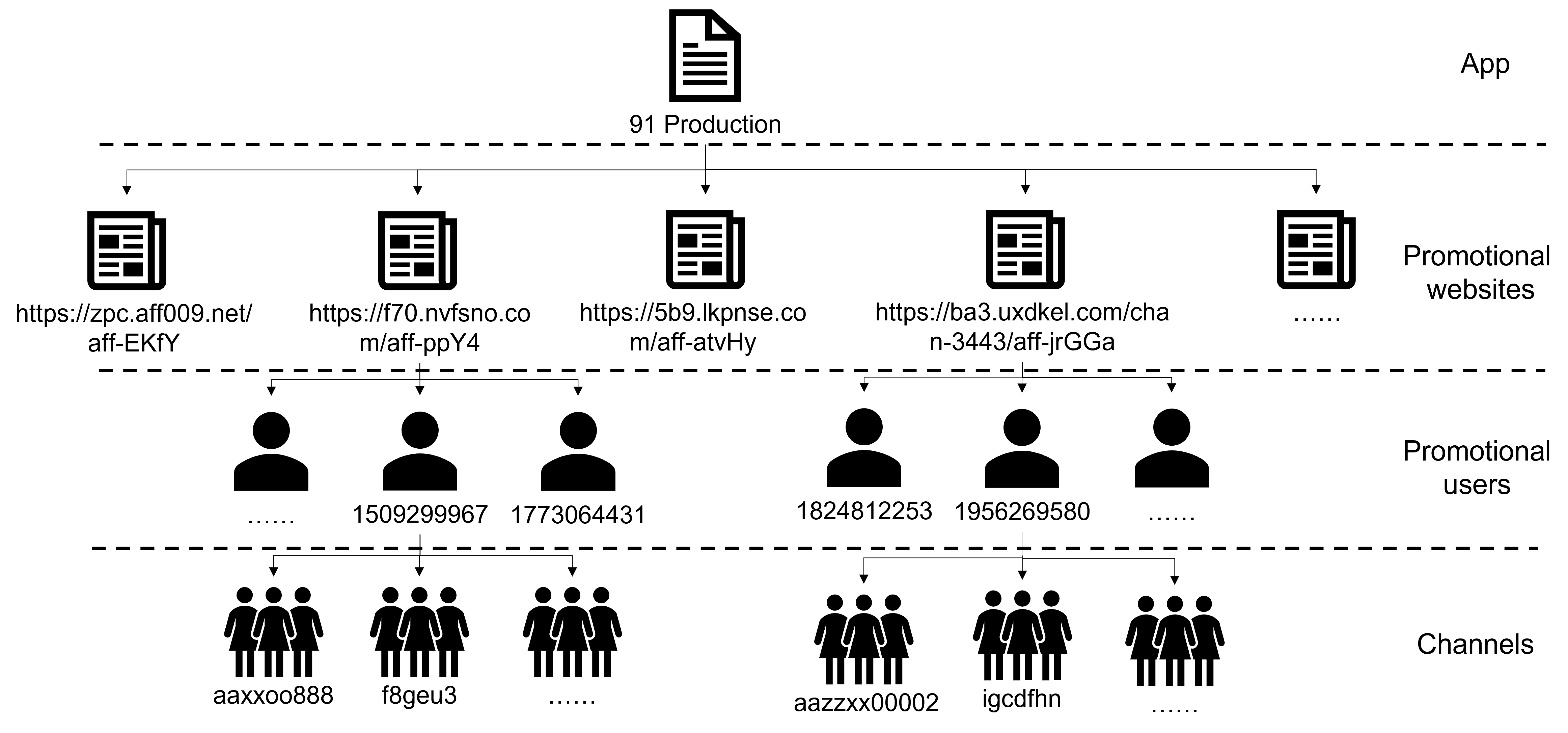}
  \caption{Promotion Model of TUApps}
  \label{promotionModel}
\end{figure*}

Our research unravels the promotional networks of TUApps by tracing their origins, 
which involve three key entities: \textit{websites} that host app files, Telegram \textit{users} who promote these websites through messages, and \textit{channels} that provide promotional space for these users. \autoref{promotionModel} illustrates this intricate promotional model of TUApps using the case of a pornography app "91 Production", demonstrating how the app is distributed across multiple websites, each promoted by different users within various channels. Our analysis of this model reveals several noteworthy observations.

\subsubsection{Promotional Websites} We identified two significant patterns regarding websites:
\begin{itemize}[leftmargin=*]
    \item \textbf{Diverse Hosting}: Promotional websites for the same TUApp tend to utilize different hosts. On average, only 11.3\% of websites promoting the same TUApp share identical secondary hosts. This diversity suggests a deliberate strategy of constantly generating new hosts for TUApp promotion, potentially as a means to evade regulatory detection and action.
    \item \textbf{URL Structural Similarities}: Despite the host diversity, we observed striking similarities in URL structures and formats used to promote the same app. For instance, as shown in \autoref{promotionModel}, websites promoting "91 Production" often incorporated the common string "aff-" at the end of their URLs. Additionally, these URLs frequently exhibited similar segmentation patterns when split by special characters (such as "/", "-", and ".").
\end{itemize}

To validate our observations on URL structural similarities, we conducted a detailed analysis of URLs promoting each TUApp in our dataset. Our methodology involved segmenting the URLs based on common delimiters (e.g., /, -, ?, \&) and excluding common segments such as top-level domains (e.g., com, net) and protocol prefixes (http, https, www). We then analyzed the remaining segments, regardless of their length. For each app, we calculated two metrics: the proportion of URLs sharing at least one unique segment with another URL promoting the same app; and the proportion of URLs having the same number of segments as the most common segment count for that app. Averaging these metrics across all apps in our dataset, our analysis revealed that: 91.46\% of URLs promoting the same app contained at least one identical segment with another URL for that app. 72.74\% of URLs maintained a consistent number of segments within each app's promotional URL set.
Further analysis of segment lengths showed that segments between 3 and 6 characters long constitute the majority, accounting for 80.67\% of all segments and 80.75\% of identical segments across URLs.
These structural similarities in URL composition, despite varying hosts, present a potential avenue for detection and mitigation strategies against TUApp promotion.

\subsubsection{Promotional Users}
Our analysis of the 7,957 users involved in TUApp promotion revealed a sophisticated and diversified approach to app marketing. These users demonstrate a high degree of versatility, rarely limiting themselves to promoting a single website or app. Instead, they engage in extensive promotion, advertising a wide range of sites associated with multiple apps across numerous channels.
On average, each promoter advocates for 5.7 distinct TUApps, utilizes 74.32 unique URLs in their promotional efforts, and operates across 13.72 different channels.

Furthermore, we identified three primary promotional strategies employed by these users:
\begin{itemize}[leftmargin=*]
    \item Top-of-Chat Advertisements: Channel administrators leverage Telegram's pinning feature to continuously display TUApp ads at the top of the chat interface to  to maximize visibility.
    \item Greeting Advertisements for Newcomers: Automated welcome messages containing ads for TUApps, set up by channel administrators, are sent to new members upon joining.
    \item Regular Advertisement Messages: Promotional users periodically post ads within channels to maintain consistent visibility of TUApps.
\end{itemize}

\subsubsection{Promotional Channels}
We conducted a comprehensive analysis of 6,137 channels involved in disseminating TUApps. Our investigation revealed that, on average, each channel hosted 8.29 promotional users, collectively promoting 7.4 apps through 293.31 promotional messages.

We sought to determine whether these channels were regular channels unrelated to the underground economy or specifically dedicated to serving underground content.
To investigate this, we employed a stratified sampling approach based on promotional message frequency.
Using quantile-based classification, we categorized channels into three groups: low density (below the first quartile, Q1), medium density (between Q1 and the third quartile, Q3), and high density (above Q3).
We then randomly selected 50 channels from each category for manual inspection. Among them, high-density channels (more than 23.42 promotional messages per day) and medium-density channels (7.69 to 23.42) were all dedicated to serving underground content. Even among low-density channels (less than 7.69), only 4 out of 50 were normal channels unrelated to underground activities.

These underground-focused channels typically revolved around specific themes such as pornography or gambling. They provided a range of services including app download links, user guides, and enticing offers promising quick money-making opportunities or free access to restricted content. Some channels also incentivized user growth by offering rewards to existing users for successfully inviting new members.
Notably, despite Telegram's efforts to alert users through "scam" channel tags, we observed a significant gap in the platform's ability to identify and flag potentially harmful channels. Out of the 6,137 promotional channels analyzed, only 17 (0.28\% of the total) were marked as scams. While not all promotional channels are necessarily scams, this low percentage suggests a considerable challenge in Telegram's capabilities to identify potentially harmful channels.

\subsection{Promotion Team Identification}
\label{promotionTeam}

We next investigate the existence of organized promotion teams among the 7,957 promotional users. Our hypothesis posited that users promoting the same website are likely members of the same team. To test this hypothesis and identify potential teams, we utilized the Louvain algorithm~\cite{blondel2008fast}, a widely recognized method for community detection in large networks.
The Louvain algorithm is renowned for its effectiveness in social network analysis and community detection within computer science research as exemplified in ~\cite{white2022characterizing,miz2020trending,moradi2021community}. We applied this algorithm to our dataset by constructing an undirected graph where nodes represent users and edges represent shared promotional websites. The weight of each edge was determined by the number of promotional sites common to the connected users.
Our constructed graph comprised 7,957 nodes (users) interconnected by 49,607 edges (shared promotional websites). The Louvain algorithm partitioned this graph into 681 distinct communities, potentially representing different promotion teams. To evaluate the quality of this community division, we employed the modularity metric, which ranges from -0.5 to 1, with higher values indicating stronger community structures. Our network exhibited a modularity of 0.955, suggesting a robust and well-defined community structure.

Our analysis of the community structure revealed a highly skewed distribution of users across clusters. The majority of communities (81\%, 552 out of 681) contained between 3 to 10 users, with a median size of 6 users. The smallest community had 3 users, while the largest contained 348 users. Notably, only 8 communities exceeded 100 users. This distribution indicates that the promotion network is characterized by numerous small teams, with a few exceptionally large communities.

Following our community detection, we examined the promotional patterns within the top 20 identified communities. This analysis revealed two distinct categories of promotional teams: those focused on promoting a single TUApp and those engaged in promoting multiple TUApps. Notably, among the top 20 identified teams, 17 were involved in promoting multiple apps. This prevalence of multi-app promotional teams strongly suggests the existence of specialized underground app promotion services operating independently from app development teams on Telegram.

To further investigate whether apps promoted by the same team were developed by the same entity, we conducted an app clustering analysis based on their similarity. 
We employed FSquaDRA2~\cite{zhauniarovich2014fsquadra} to calculate similarity scores between each pair of apps, forming clusters based on these scores.
While initially designed as an app repackaging detection tool, FSquaDRA2 has proven versatile in broader app similarity analysis and clustering tasks. Its effectiveness in these areas has been demonstrated in several recent studies~\cite{gao2021demystifying,li2019identifying,hu2019dating}. The tool's ability to assess app similarity makes it well-suited for our purpose of identifying apps with shared development origins or significant structural similarities.
Following established research practices, we set a clustering threshold of 0.8, grouping apps with similarity scores exceeding this value. Additionally, we clustered apps with identical digital signatures, as this strongly indicates a common development origin.
This analysis resulted in the distribution of 557 Android apps into 41 clusters, with 32 clusters comprising multiple apps. \autoref{appCluster} presents an overview of the five largest clusters, detailing the number of apps per cluster, signature quantities, and observed patterns in package and MainActivity naming.

\begin{table}[htb]
  \centering
  \small
  \caption{Top 5 App Clusters Information}
  \begin{tabular}{p{1cm}p{1cm}p{1cm}c}
    \hline
App Count & Signature Count & Package Name & MainActivity Name          \\
    \hline
27        & 3               & live.*.*     & com.tbone.*.MainActivity   \\
15        & 2               & tv.*.*       & com.example.*.MainActivity \\
13        & 1               & net.*.*      & com.tast.*.MainActivity    \\
13        & 2               & vip.*.*      & com.awjq.*.MainActivity    \\
8         & 1               & gov.*.*      & com.pron.*.MainActivity    \\
    \hline
  \end{tabular}
  \label{appCluster}
\end{table}

We then mapped the promotion teams to these app clusters, revealing a many-to-many relationship between promotion teams and development teams. Among the top 20 promotion teams, 7 exclusively promoted apps from a single development team, while the remaining 13 promoted apps from multiple development teams. This finding not only confirms the existence of organized app promotion teams operating independently of development teams but also aligns with our observations from search bot interactions, where promotion teams proactively reached out and offered their services to Telegram users.

These results highlight the complex ecosystem of underground app promotion on Telegram, suggesting a sophisticated and adaptable underground economy capable of efficiently distributing and promoting illicit apps across the platform.

\subsection{Promotional Website Analysis}
To gain insights into the operational characteristics of websites promoting TUApps, we conducted an analysis focusing on their survival time and infrastructure details.

\subsubsection{Website Operational Duration}
We assessed the operational lifespan of promotional websites using Python's requests library~\cite{request} to perform daily status checks. Websites returning a 200 HTTP status code were considered active, while any other response indicated inactivity. 
Our analysis of website longevity revealed several key findings:
\begin{itemize}
\item Overall, promotional websites for underground apps exhibited short lifespans, with an average operational duration of just 24 days, indicating their transient nature.
\item Websites promoting pornography apps demonstrated notably short lifespans, with an average operational duration of 18.5 days. Nearly 90\% of these sites ceased operations within a month, suggesting a high-turnover, high-risk environment.
\item Gambling app websites demonstrated relatively greater resilience, maintaining an average lifespan of 33.2 days. Some outliers in this category showed exceptional longevity, remaining active for up to six months, potentially indicating a more stable and lucrative market.
\end{itemize}
The generally short lifespan of these promotional websites underscores the challenges faced by both operators in maintaining their online presence and by authorities in tracking and regulating these rapidly shifting platforms.

\subsubsection{Infrastructure and Registration Analysis}
\label{InfrastructureandRegistration}
We further investigated the network infrastructure and registration details of the promotional websites. We employed the \textit{ipinfo.io}~\cite{IPinfo} service to identify the Autonomous System Numbers (ASNs) corresponding to the websites' IP addresses. Domain registrar information was obtained through \textit{whois}~\cite{whois} searches.
Key findings include:
\begin{itemize}
\item As shown in \autoref{asnOfWeb}, a significant number of ASNs were predominantly hosted in the United States, China, and Malaysia.
\item \autoref{RegistrarOfWeb} illustrates the domain registrar landscape, with \textit{GoDaddy} and \textit{Name.com} collectively managing more domains than the combined total of the next eight competitors.
\end{itemize}
This concentration of hosting and registration services suggests potential focal points for future regulatory or investigative efforts in combating the distribution of underground apps.

\begin{table}[htb]
\centering
\small
\caption{Top 10 ASNs of Promotion Websites}
\begin{tabular}{cccc}
\hline
ASN & ASN Description & Count & Country\\
\hline
AS13335 & Cloudflare, Inc. & 898 & US \\
AS55720 & Gigabit Hosting Sdn Bhd & 501 & MY \\
AS16509 & Amazon.com, Inc. & 494 & US \\
AS36351 & SoftLayer Technologies Inc. & 458 & US \\
AS45090 & Shenzhen Tencent Computer Systems Company Limited & 253 & CN \\
AS140227 & Hong Kong Communications International Co., Limited & 215 & CN \\
AS134548 & DXTL Tseung Kwan O Service & 211 & CN \\
AS37963 & Hangzhou Alibaba Advertising Co.,Ltd. & 205 & CN \\
AS4847 & China Networks Inter-Exchange & 104 & CN \\
AS45062 & Netease-Network & 104 & CN \\
\hline
\end{tabular}
\label{asnOfWeb}
\end{table}

\begin{table}[htb]
  \centering
  \small
  \caption{Top 10 Registrars of Promotion Websites}
  \begin{tabular}{p{7cm}c}
    \hline
    Registrar & Count \\
    \hline
    GoDaddy.com, LLC & 1118 \\
    Name.com, Inc. & 745 \\
    Gname.com Pte. Ltd. & 321 \\
    eName Technology Co.,Ltd. & 182 \\
    Alibaba Cloud Computing (Beijing) Co., Ltd. & 152 \\
    GANDI SAS & 151 \\
    Alibaba Cloud Computing Co., Ltd. (Wanwang) & 84 \\
    Alibaba Cloud Computing Ltd. d/b/a HiChina & 76 \\
    Go Daddy, LLC & 65 \\
    NameSilo, LL & 58 \\
    \hline
  \end{tabular}
  \label{RegistrarOfWeb}
\end{table}

\answer{2}{The promotion of underground apps on Telegram reveals a sophisticated ecosystem characterized by intricate relationships among apps, websites, users, and channels. 
Our analysis uncovered that promotional websites employ diverse hosting strategies while maintaining structural similarities in URLs, potentially offering avenues for detection and mitigation. 
The vast majority of channels involved in app promotion are dedicated to underground content, with only 0.28\% flagged as scams by Telegram, indicating significant gaps in platform-level detection.
We identified organized promotion teams, some specializing in single apps while others manage multiple apps, showcasing a mature and adaptable underground economy.
Promotional websites for TUApps demonstrate remarkably short operational periods of merely 24 days on average.
The network infrastructure supporting these promotional activities is primarily concentrated in the United States and China. The complex promotional landscape underscores the challenges in regulating and mitigating the distribution of underground apps on encrypted messaging platforms.}
\section{Characterizing TUApps}
\label{sec:rq3sec}
This section provides an in-depth analysis of the characteristics of TUApps, focusing on their development features and behavioral patterns.

\subsection{App Development Characteristics}
We investigate the key development characteristics of TUApps, exploring three critical aspects: development frameworks, third-party library integration, and app signatures. This analysis provides insights into the technical choices and strategies adopted by underground app developers.

\subsubsection{Development Framework Analysis}
To understand the technological foundation of TUApps, we examined the development frameworks used in their creation. Following a methodology similar to~\cite{hong2022analyzing}, we analyzed 35 popular mobile app development frameworks, focusing on their distinctive features such as dynamic link libraries (.so) and header files (.h). We then performed matching within each app to identify the frameworks used.
Our analysis revealed a significant preference for cross-platform development frameworks among TUApp developers. The four most commonly used frameworks were:
Flutter (39.0\%),
React Native (6.6\%),
Cordova (5.6\%),
and Unity3D (3.9\%).
Notably, Flutter~\cite{flutter} emerged as the dominant choice, likely due to its ease of use and rapid development capabilities. This preference for Flutter in TUApps (39.0\%) stands in stark contrast to its usage in regular apps, which according to AppBrain~\cite{appBrain}, is only 4.88\%.
The prevalence of cross-platform frameworks in TUApp development presents unique challenges for app analysis. Traditional app analysis tools, including static analysis frameworks like FlowDroid~\cite{arzt2014flowdroid} and dynamic analysis tools such as DroidBot~\cite{li2017droidbot}, are primarily designed for native apps and offer limited support for cross-platform apps. This discrepancy highlights a critical need for advanced analytical tools capable of effectively examining cross-platform apps, particularly in the context of identifying and analyzing underground apps.

\subsubsection{Third-Party Library Analysis} 
We utilized LibRadar~\cite{ma2016libradar} to analyze and catalog the embedded third-party resources in each app. Our investigation revealed that TUApps incorporate, on average, approximately 14.3 external libraries.
While many of these libraries, such as OKHttp and Fastjson, are commonly used in regular app development, we observed some notable anomalies. Surprisingly, the most frequently incorporated library was ZXing~\cite{zxing} (71.1\%, used in 396/557 apps), a tool for generating and decoding QR codes. This prevalence suggests a significant reliance on QR code functionality within TUApps, potentially for sharing app links or facilitating transactions.
Additionally, we found a high frequency of payment-related libraries, with AliPay being particularly prominent (64.1\%). This frequent integration of payment capabilities indicates that many TUApps are designed with monetization as a key feature, likely facilitating direct financial transactions within the apps.

\subsubsection{Signature Analysis}

We used Androguard~\cite{androguard} and \seqsplit{plistlib}~\cite{plistlib} to extract the signatures of Android and iOS apps.
A closer inspection of the certifying authorities revealed stark distinctions between the iOS and Android ecosystems, as shown in \autoref{signAuth}.
For iOS apps, the certifying entities turned out to be corporations that could be identified through search engines, suggesting the possible misappropriation of signature validation. Conversely, the credentials identifying Android apps' certifying authorities were often ambiguous, and app developers seemed to deliberately obscure their identities.

\begin{table}[htb]
\centering
\small
\caption{Top 5 Signatures Authorities for iOS and Android}
\begin{tabular}{cc}
\hline
iOS & Android \\
\hline
Nan Chang Wo Ai Wo Jia Technology CO.Ltd & dd \\
Efka - IIektronnikons Ethnikos Foreas Koinokis Asfalisis & gjlc \\
AVIATION INFORMATION AND TELECOMMUNICTIONS JSC & momo13 \\
Baretch (International) Information Network Ltd & AW \\
EARLY MAKERS GROUP & baichaocan2022 \\
\hline
\end{tabular}
\label{signAuth}
\end{table}

\subsection{App Behavior Characteristics}

This section investigates the behavioral patterns of TUApps to identify potential security concerns. Our analysis encompasses app network traffic, maliciousness assessment, and payment method examination.

\subsubsection{Traffic Analysis}
\label{networkAnalysis}

To analyze app network traffic, we conducted automated testing of all apps while capturing their network communications. We employed BurpSuite~\cite{BurpSuite}, a widely-used cybersecurity tool, for intercepting and analyzing network traffic.
Our testing methodology varied by platform.
For Android apps, we installed apps on a Pixel 3 device via ADB and used FastBot~\cite{cai2020fastbot} for automated testing. 
For iOS apps, We used ideviceinstaller to install IPA files to an iPhone 8 and employed macaca~\cite{macaca} for automated testing.
For web clips, we extracted and tested URLs directly in the Safari browser.
Each app test ran for five minutes, during which we captured not only network traffic, but also UI screenshots after each step of automated testing for the subsequent maliciousness analysis.

Our examination of the network traffic associated with underground apps revealed patterns in domain usage and server affiliations. \autoref{topleveldomain} illustrates the most frequently utilized top-level domains (TLDs) in this traffic.
While traditional generic top-level domains (gTLDs) such as .com and .org dominate, we observed a notable trend towards newer gTLDs like .vip and .app, as well as country code top-level domains (ccTLDs) such as .me. The preference for these alternative TLDs can be attributed to their cost-effectiveness and relatively relaxed management policies, making them particularly attractive for underground app operations.

\begin{table}[htb]
  \centering
  \small
  \caption{Top-level Domain Distribution}
  \begin{tabular}{cccc}
    \hline
    Top-level Domain & Category & \# count & \% Percent \\
    \hline
    .com & gTLD & 5,472 & 62.40 \\
    .me & ccTLD & 704 & 8.02 \\
    .vip & New gTLD & 576 & 6.56 \\
    .org & gTLD & 544 & 6.20 \\
    .app & New gTLD & 480 & 5.47 \\
    .top & New gTLD & 288 & 3.64 \\
    .shop & New gTLD & 160 & 1.82 \\
    .xyz & New gTLD & 160 & 1.82 \\
    \hline
  \end{tabular}
  \label{topleveldomain}
\end{table}

Extending our analysis beyond TLDs, we examined server affiliations within the network traffic, focusing on ASNs and domain registrars. We employed methodologies similar to those used in analyzing promotional websites (\S\ref{InfrastructureandRegistration}). This investigation yielded results similar to those observed in promotional websites: ASNs were predominantly hosted in the United States and China, with GoDaddy emerging as the most prevalent domain registrar.
These parallel findings underscore the interconnected nature of the underground app ecosystem, where similar infrastructure choices are made across app operation and promotion. This consistency may offer insights for identifying and potentially mitigating the spread of underground apps.

\subsubsection{Maliciousness Analysis}
Our maliciousness analysis began by scanning each TUApp using VirusTotal~\cite{VirusTotal}, a leading online platform that aggregates antivirus results from over 60 security vendors. 
Given a sample, VirusTotal aggregates the detection labels from various antivirus engines and reports how many of them classify the sample as ``malicious''. Generally, this number serves as an indicator to assess the level of a sample's maliciousness~\cite{wang2023re}, i.e., the larger the number, the greater the maliciousness.
As a result, 142 apps (representing 25.5\% of the dataset) were flagged as malicious by VirusTotal engines.
The number of engines identifying these apps as malicious ranged from 4 to 26, with an average of 15.04 engines.
Given that many researchers consider detection by more than 4 engines sufficient to deem a sample malicious~\cite{TESSERACT,shen2022large,zhu2020measuring}, this finding suggests that TUApps exhibit a significant level of maliciousness. 

To further classify these malicious apps, we employed AVClass2~\cite{sebastian2020avclass2}, a widely-used malware family labeling tool. The apps were categorized into various families. \autoref{Top 5 Malicious Famliy Names} presents the top 5 malware families identified, detailing their prevalence and typical behaviors. 'Jiagu' was the most prevalent, identified in 32 apps (22.5\%), and is known for packing app code to evade detection. This was followed by 'Piom', a Trojan that spies on user behavior, and 'Softpulse', which displays numerous unwanted ads. This family distribution highlights the diverse malicious behaviors present in the TUApps dataset.

\begin{table}[htb]
  \centering
  \small
  \caption{Top 5 Malware Families}
  \begin{tabular}{cccc}
    \hline
    Famliy Name & \# Count & \% Percent & Famliy Description\\
    \hline
        jiagu & 32 & 22.5 & Packs app code to evade detection\\
        piom & 31 & 21.8 & Trojan that spies on user behavior~\cite{piom1,piom2}\\
        softpulse & 25 & 17.6 & Display numerous unwanted ads~\cite{softpulse}\\
        boogr & 19 & 13.4 & Downloads other malicious files~\cite{boogr}\\
        mobtes & 17 & 11.9 & Performs stealthy destruction actions~\cite{Mobtes}\\
    \hline
  \end{tabular}
  \label{Top 5 Malicious Famliy Names}
\end{table}

Digging deeper into their behavior, our manual inspection of the screenshots captured during automated testing revealed two key findings:
\begin{itemize}
    \item Deceptive Practices: Many apps employed misleading icons and names, mimicking popular legitimate apps to lure users and disguise their true functionality. \autoref{name} illustrates examples of such deceptive naming practices.
    \item Superuser Requests: We identified 47 pornography apps requested superuser rights at startup, with 24 apps checking for root status. These apps would terminate if superuser access was denied on rooted devices. Notably, this behavior was absent in gambling apps.
\end{itemize}

\begin{table}[htb]
  \centering
  \small
  \caption{Underground Apps Mimicking Popular Apps}
  \begin{tabular}{cc}
    \hline
    Underground App & Mimicked App\\
    \hline
    DiDi & DiDi \\
    Pilipili & Bilibili \\
    JinDong & JingDong \\
    Kuaishou Express & Kuaishou \\
    Adult Bilibili & Bilibili \\
    \hline
  \end{tabular}
  \label{name}
\end{table}

These findings underscore the potential security risks associated with TUApps and highlight the importance of robust security measures in identifying and mitigating threats within the underground app ecosystem.

\subsubsection{Payment Behavior}
To analyze payment behaviors in TUApps, we conducted a manual examination of a random sample of 100 apps. Among these, 83 offered functionalities for purchasing memberships or recharging accounts. We simulated the order placement process in these 83 apps, halting just before the actual payment step.
Our investigation uncovered a sophisticated payment system employed by 69 of the examined apps.
These apps redirected users to external web browsers to complete transactions through fourth-party payment platforms~\cite{gao2021demystifying}. 
These fourth-party platforms serve as intermediaries, routing payment requests from the apps to well-known third-party services such as WeChat Pay or Alipay. 
\autoref{pay} provides examples of five fourth-party platforms we identified during our analysis.

\begin{table}[htb]
  \centering
  \small
  \caption{Examples of Fourth-party Payment Services Identified}
  \begin{tabular}{ccc}
    \hline
    Platform & Alipay & WeChat Pay \\
    \hline
    http://ngxkdeja.xingyuwl.xyz & Yes & No \\
    http://www.weepay.top & No & Yes \\
    https://wapcashier.ysepay.com & Yes & No \\
    http://pj888.005c.com & Yes & No \\
    http://xin.xinma.work & No & Yes \\
    \hline
  \end{tabular}
  \label{pay}
\end{table}

A key feature of this payment system is its use of dynamic virtual accounts for each transaction. This practice effectively obscures the ultimate destination of funds, creating a maze of financial transfers that is challenging to trace.
This payment mechanism offers a high degree of anonymity through its multi-layered forwarding and complex account utilization. Such structures are also susceptible to exploitation for money laundering activities~\cite{tiwari2024trade}. Potential money launderers could leverage these intricate payment processes to convert illicit funds into seemingly legitimate transactions. This complexity adds another layer of challenge to regulatory and law enforcement efforts in monitoring and controlling the financial aspects of the underground app ecosystem.

\answer{3}{Our analysis of TUApps reveals distinctive characteristics in both app development and behavior. On the development front, TUApps predominantly favor Flutter for ease of development, heavily integrate QR code and payment libraries, and exhibit significant differences in app signatures between Android and iOS platforms. Behaviorally, TUApps employ a mix of traditional gTLDs and new TLDs, with their network infrastructure primarily concentrated in the United States and China. Notably, one-fourth of the analyzed TUApps are classified as malware, displaying a high degree of maliciousness. Furthermore, these apps frequently utilize fourth-party payment platforms, likely as a strategy to evade regulatory scrutiny. These findings underscore the sophisticated nature of TUApps, combining advanced development techniques with potentially malicious behaviors and evasive financial practices.}
\section{Discussion}
\subsection{Implications and Potential Mitigations}
\label{sec:miti}
Our comprehensive study of the underground mobile app ecosystem on Telegram reveals several critical implications and potential avenues for mitigation. These insights are valuable for platform regulators, app market operators, law enforcement agencies, and cybersecurity professionals.

Firstly, our analysis demonstrates that despite Telegram's efforts to warn users through scam tagging, current measures have failed to effectively regulate the presence of potentially harmful content on the platform. The encrypted communication and anonymity offered by Telegram, while valuable features, inadvertently provide a haven for illicit activities. This situation is particularly concerning given our finding that a quarter of TUApps exhibit maliciousness. To address this, we propose leveraging our findings to enhance Telegram's detection capabilities. By developing more sophisticated content analysis tools capable of identifying patterns in promotional messages, channel behaviors, and URL structures associated with underground apps, Telegram could significantly improve its ability to detect potentially harmful content. This could be coupled with an in-app warning system that alerts users before they interact with suspicious content or click on external links, thereby reducing the risk of harm and enhancing overall platform safety and user trust.

Secondly, although iOS's restrictions on sideloading IPA files serve as a deterrent to the distribution of underground apps, the exploitation of iOS system features—such as TestFlight, Enterprise Signing, and Web Clips—for the distribution of TUApps demands attention from mobile system platforms. It is crucial for these platforms to further enhance regulatory measures to address these vulnerabilities or loopholes to prevent the misuse of their system features for distributing underground apps. Moreover, the discovery of underground apps temporarily available on the official Apple App Store underscores the importance of continually refining app review processes.

Thirdly, the prevalence of fourth-party payment processors in underground apps raises concerns about potential money laundering activities. The complex web of transactions involving multiple layers of intermediaries can obscure the origin and destination of funds, making it challenging for authorities to trace illicit financial flows. We suggest collaborating with financial institutions to develop more advanced transaction monitoring systems capable of identifying suspicious patterns associated with these layered payment structures.

Our findings not only shed light on the sophisticated nature of this ecosystem but also provide concrete pathways for enhancing detection, prevention, and regulatory measures. By implementing these recommendations, stakeholders can leverage the insights from our research to develop more effective strategies for combating the proliferation of underground apps.

\subsection{Limitation}

Our research encountered several limitations, primarily in data collection and the scope of our analysis. We employed multiple methods to source channels promoting underground apps, including bot-based searches, collection of top regular channels, and snowballing techniques. However, we observed significant disparities in the effectiveness of search bots between English and Chinese channels, with Chinese search bots yielding thousands of relevant channels in response to keyword queries, while English bots returned only dozens.
Despite our intention to maintain a language-agnostic approach in data collection, utilizing both English and Chinese keywords (as detailed in \S~\ref{keywords}), our dataset predominantly captured underground apps and promotional activities targeting Chinese-speaking users. This skew likely reflects China's stringent regulatory environment, particularly concerning pornography and gambling, which may drive such activities to seek refuge in discreet platforms like Telegram. In contrast, regions with more relaxed regulations might allow similar content to be directly accessible via websites or social media, potentially reducing the demand for dedicated apps.

While our study offers valuable insights into the Chinese-language underground app ecosystem within Telegram, this predominance may limit the generalizability of our findings to other linguistic or geographic contexts. Nonetheless, we believe that the patterns and dynamics observed can serve as a foundation for understanding the challenges associated with underground app distribution on encrypted messaging platforms, especially in environments with strict content regulations.
Researchers interested in underground apps targeting users of other languages could apply our approach, leveraging language-specific search bots and keywords to identify relevant channels and uncover underground app ecosystems within their linguistic contexts. 

Furthermore, the reliance on public channels for data collection may overlook underground apps promoted through private channels and one-on-one communications, potentially underestimating the scale and diversity of the underground economy. Additionally, the dynamic nature of the underground app market,  with rapidly emerging new apps and promotion strategies, may limit the long-term applicability of the findings. Our focus on Telegram means that the findings may not be directly applicable to other platforms, such as Discord~\cite{telecloudsek}, which warrant separate investigation in future work.

\section{Related Work}
\subsection{Telegram's Underground Ecosystem}
Previous studies on Telegram's underground ecosystem primarily focuses on the conversation channels and their thematic content. Morgia et al.~\cite{la2021uncovering} delved into Telegram's dark side by examining channels marked as fake or scam, exploring the nature and operations of clones and conspiracy movements. Further extending their research~\cite{la2023tgdataset}, they amassed a collection of more than 100,000 Telegram channels to study the spread of illicit content, such as pornography, fraudulent credit card activities, violent content, hacking, and white supremacist ideologies. They also studied the dissemination of misinformation through Telegram channels, alongside proposing detection techniques for these deceptive channels~\cite{la2023sa}. Vincenzo et al.~\cite{imperati2023conspiracy} undertook a thorough analysis of conspiracy theories circulating on Telegram, focusing on the identification of channels promoting such theories and their revenue-generating strategies. In contrast to previous research that focused on channels and thematic content, our study concentrates on mobile underground apps in Telegram.

\subsection{Analysis of Underground Apps}
Existing research on underground apps primarily revolves around their acquisition, identification, and analysis within the underground app ecosystem.

Regarding the acquisition of underground apps, the strategies generally involve either cooperation with authoritative institutions or analyzing promotional websites for these apps. Chen et al.~\cite{chen2023deuedroid}, Hu et al.~\cite{hu2022measurement} and Hong et al.~\cite{hong2022analyzing} succeeded in obtaining underground app samples through collaborating with authoritative institutions. Han et al.~\cite{han2023measurement} and Gao et al.~\cite{gao2021demystifying} acquired gambling-related domain names from authoritative institutions and crawled corresponding websites to obtain gambling apps. Through traffic analysis, Chen et al.~\cite{chen2024underground} identified underground online portals and crawled underground apps from them.

To identify underground apps, Chen et al.~\cite{chen2023deuedroid} employed the similarity of UI transition graphs to detect underground apps. Zhao et al.~\cite{zhao2024no} identified mask apps—malicious apps disguised as legitimate apps—on the Apple App Store, leveraging features including discrepancies between app descriptions and user reviews, app recommendation relationships, and code similarities.

Regarding the ecosystem analysis of underground apps, Han et al.~\cite{han2023measurement} explored the ecosystem of illegal Android gambling apps by examining the relationship between gambling apps and advertising campaigns on upper-level pages. From a different angle, Gao et al.~\cite{gao2021demystifying} focused on the ecosystem of illegal mobile gambling apps, analyzing their distribution methods, key characteristics, and identification methods. Hu et al.~\cite{hu2022measurement} studied the usage of underground Web Clips on the iOS platform from the perspectives of creators, distributors, and operators. Hong et al.~\cite{hong2022analyzing} analyzed the chain of app gambling scams, starting from fraud alerts, and examined the development frameworks, permissions, compatibility, and network infrastructure of fraudulent apps. Chen et al.~\cite{chen2021lifting} revealed the structure of the underground app ecosystem and the operation of app builders by studying four participating entities – developers, proxies, operators, and harvesters – along with their workflows. Hu et al.~\cite{hu2019dating} conducted an in-depth study of the ecosystem of fraudulent dating apps, analyzing the producers, publishers, and distribution networks, as well as identifying such apps. 
Our work complements these studies by specifically focusing on the underground app ecosystem in Telegram.
\section{Conclusion}
This study presents the first comprehensive analysis of the underground mobile app ecosystem on Telegram, illuminating a previously understudied facet of digital underground economies. By constructing and analyzing a novel dataset, we reveal a sophisticated network of underground apps nominally reaching 1\% of Telegram's user base. Our research uncovered several critical findings, including the misuse of iOS features for app distribution, significant gaps in Telegram's content moderation capabilities, and the alarming prevalence of malicious behaviors in these underground apps. The widespread use of fourth-party payment platforms further complicates regulatory oversight. Our work underscores the urgent need for enhanced regulatory measures and detection mechanisms on encrypted platforms. Our findings provide valuable insights for platform regulators, app market operators, law enforcement agencies, and cybersecurity professionals in their efforts to combat underground app proliferation and protect users from associated risks.

\section*{Acknowledgment}

We sincerely thank our shepherd Prof. Andrea Marin (Università Ca' Foscari di Venezia) and all the anonymous reviewers for their valuable suggestions and comments to improve this paper. 
This work was supported by the Key R\&D Program of Hubei Province~(2023BAB017, 2023BAB079), the National NSF of China (grants No.62072046, 62302181), the Xiaomi Young Talents Program, the HUSTCSE-FiberHome Joint Research Center for Network Security, and the Beijing University of Posts and Telecommunications 2023 Education and Teaching Reform Project (2023YB37).

\bibliographystyle{ACM-Reference-Format}
\bibliography{base}

\newpage
\appendix
\section{Appendix}
\subsection*{Ethics and Availability}
\label{ava}
Our research involved the collection and analysis of a large-scale dataset from public Telegram channels. While these channels are publicly accessible, we acknowledge that the users participating in these channels may not have explicitly consented to the collection and use of their data for research purposes. To address these concerns, we have taken measures to ensure the responsible handling of the collected data:
\begin{itemize}
\item We have implemented secure data storage practices. The data is stored on encrypted drives with restricted access limited to authorized researchers directly involved in this study.
\item In reporting our findings, we have presented the results in an aggregated and anonymized manner, avoiding the inclusion of personally identifiable information.
\end{itemize}
Regarding availability, our code and dataset are available at \url{https://github.com/security-pride/TUApps}. Specifically:
\begin{itemize}
\item The underground apps discovered in our research will be made publicly available, as they are essential for understanding and combating illicit activities within the mobile app ecosystem.
\item To protect user privacy, the dataset containing user messages will be available privately upon request from researchers who agree to adhere to strict ethical guidelines and maintain data confidentiality.
\item The code used for data collection and analysis will also be available privately upon request to prevent potential misuse and ensure that future research follows appropriate ethical standards. 
\end{itemize}
We believe that this approach balances the need for transparency and reproducibility in research with the importance of protecting user privacy and preventing unethical data collection practices.

\end{document}